\documentclass[aip,reprint]{revtex4-1}
\usepackage{graphicx,epstopdf,bm}
\usepackage{subfigure}
\usepackage{soul}
\usepackage{hyperref}
\hypersetup{colorlinks=true,linkcolor=blue,citecolor=blue,urlcolor=blue}

\usepackage{color}

\newcommand{\Fkt}[1]{\,\mathsf {#1}}
\ifx\Tr\renewcommand{\Tr}{\Fkt{Tr}}
\else\newcommand{\Tr}{\Fkt{Tr}}
\fi

\begin{document}


\title{Coherent Control of Ultrafast Bond Making and Subsequent Molecular Dynamics: 
Demonstration of Final-State Branching Ratio Control}



\author{Liat Levin}
\affiliation{The Shirlee Jacobs Femtosecond Laser Research Laboratory,
  Schulich Faculty of Chemistry,
  Technion-Israel Institute of Technology, Haifa 32000, Israel}

\author{Daniel M. Reich}
\affiliation{Dahlem Center of Complex Quantum Systems and Department of Physics, Freie Universit\"at Berlin, Berlin, Germany}

\author{Moran Geva}
\affiliation{The Shirlee Jacobs Femtosecond Laser Research Laboratory,
  Schulich Faculty of Chemistry, Technion-Israel Institute of
  Technology, Haifa 32000, Israel}

\author{Ronnie Kosloff}
\affiliation{Fritz Haber Research Centre and The
    Department of Physical Chemistry, Hebrew University, Jerusalem
    91904, Israel}

\author{Christiane P. Koch}
\affiliation{Dahlem Center of Complex Quantum Systems and Department of Physics, Freie Universit\"at Berlin, Berlin, Germany}

\author{Zohar Amitay}
\thanks{E-mail: amitayz@technion.ac.il}
\affiliation{The Shirlee Jacobs Femtosecond Laser Research Laboratory, Schulich Faculty of Chemistry, 
Technion-Israel Institute of Technology, Haifa 32000, Israel}


\date{\today}

\begin{abstract}
Quantum coherent control of ultrafast bond making 
and the subsequent molecular dynamics 
is crucial for the realization of a new photochemistry,  where a shaped laser field is actively 
driving the chemical system in a coherent way from the thermal initial state of the reactants to the final state of the desired products.
We demonstrate here coherent control over the relative yields 
of Mg$_2$ molecules that  are generated via photoassociation and subsequently photodriven into different groups of final states. 
The strong-field process involves non-resonant multiphoton femtosecond photoassociation of a pair of thermally hot magnesium atoms 
into a bound Mg$_{2}$ molecule and subsequent molecular dynamics on electronically excited states.
The branching-ratio control is achieved with linearly chirped laser pulses,
utilizing the different chirp dependence that various groups of final molecular states display for their post-pulse population.
Our study establishes  the feasibility of high degree coherent control over quantum molecular dynamics 
that is initiated by femtosecond photoassociation of thermal atoms.
\end{abstract}

\pacs  {42.65.Re, 82.50.Nd, 82.53.Eb, 82.53.Kp}

\maketitle 


A driving force for the field of quantum coherent control 
has been the idea of generating a new type of photochemistry~\cite{tannor1,RonnieDancing89,RiceBook,ShapiroBook} 
in which a shaped laser field actively drives the irradiated quantum system in a coherent way 
from the initial state of the reactants to the final state of the desired products. 
This is conceptually different from the common photochemistry approach, 
where the reactants are individually photo-excited and then react without further interaction with external fields.
The key tool for the coherent control of binary reactions
are shaped femtosecond laser pulses.\cite{WeinerRevSciInst00}      
Typically, coherent control of a binary reaction proceeds in several steps. 
First, ultrafast coherent bond making (i.e., free-to-bound femtosecond photoassociation) 
is photo-induced between the free colliding reactants, and an excited bound molecule is generated.
Then, subsequent photo-control over the molecular dynamics directs the system into target intermediate states.
Last, these intermediate states serve as a platform for further coherent photo-control in order to selectively break 
bonds (i.e., femtosecond photodissociation) for generating the desired products.

Despite the significance, there is still no realization of this full scheme of binary reaction control from reactants to products. 
A major reason is the lack of successful coherent control of ultrafast bond making and the subsequent molecular dynamics. 
For binary gas phase photo-reactions where the reaction mechanism is most easily unraveled, 
such control has been demonstrated so far only in a few studies we have conducted.~\cite{RybakPRL11,AmaranJCP13,LevinPRL15,LevinJPhysB15}
Control of photoassociation was also demonstrated at ultralow temperatures with a timescale five orders of magnitude 
longer.~\cite{carini2015,ciamei2017efficient,blasing2018,kallush2017}
Quantum control of bond formation was also observed in laser-induced catalytic surface reactions~\cite{nuernberger2010,nuernberger2012} without, however, full insight into the reaction mechanism.
The key element is distillation of an entangled pair from a thermal ensemble.
In strong contrast, many studies have successfully demonstrated photodissociation control of bound molecules, 
controlling both the total yield of fragments as well as their branching ratio to different channels.~\cite{GordonARPC97,BrixnerCPC03,DantusCR04,WollenhauptARPC05,SFB450book,LevisScience2001,RabitzScience2000,toth2020,iwamoto2020strong}

The strong-field process in our previous photoassociation investigations~\cite{RybakPRL11,AmaranJCP13,LevinPRL15,LevinJPhysB15}
involves non-resonant multiphoton femtosecond photoassociation of a pair of thermally hot magnesium atoms 
into a bound excited Mg$_{2}$ molecule and subsequent molecular dynamics on electronically excited states.
At the first stage, we have demonstrated the generation of purity and rovibrational coherence in the ensemble of photoassociated 
Mg$_{2}$ molecules via the mechanism of Franck-Condon filtering.\cite{RybakPRL11,AmaranJCP13}
Then, we have utilized the generated purity and coherence to demonstrate coherent control over the resulting 
yield of molecules populating  a single group of final states.~\cite{LevinPRL15,LevinJPhysB15}
The molecular yield was enhanced with linearly-chirped femtosecond pulses having a positive chirp, 
reaching a maximum with a specific optimal chirp value,
as well as with positively chirped pulses split into sub-pulses with a temporal structure that fits 
the photo-induced coherent vibrational dynamics.
The different excitation and control mechanisms have been qualitatively explained by our \textit{ab initio} calculations.

In view of controlling the full scheme of a binary reaction from reactants to products, the main limitation of our 
previously demonstrated bond-making control has been 
the control target, i.e., 
the overall molecular yield.\cite{RybakPRL11,AmaranJCP13,LevinPRL15,LevinJPhysB15} 
Here, we  extend our approach to coherently controlling the relative yields 
of molecules that are generated via photoassociation and subsequently photo-driven into multiple groups of final states. 
The investigated ultrafast strong-field process is, as before, photoassociation of thermal magnesium atoms
followed by excited-state dynamics of the generated Mg$_{2}$ molecules.
The demonstrated branching-ratio control is achieved with linearly-chirped pulses,
utilizing the different chirp dependence that different groups of final molecular states display for their post-pulse population.
Our results establish the feasibility of high degree coherent control over quantum molecular dynamics 
that is initiated by femtosecond photoassociation of thermal atoms.

\begin{figure}[tbp]
\includegraphics[width=0.475\textwidth]{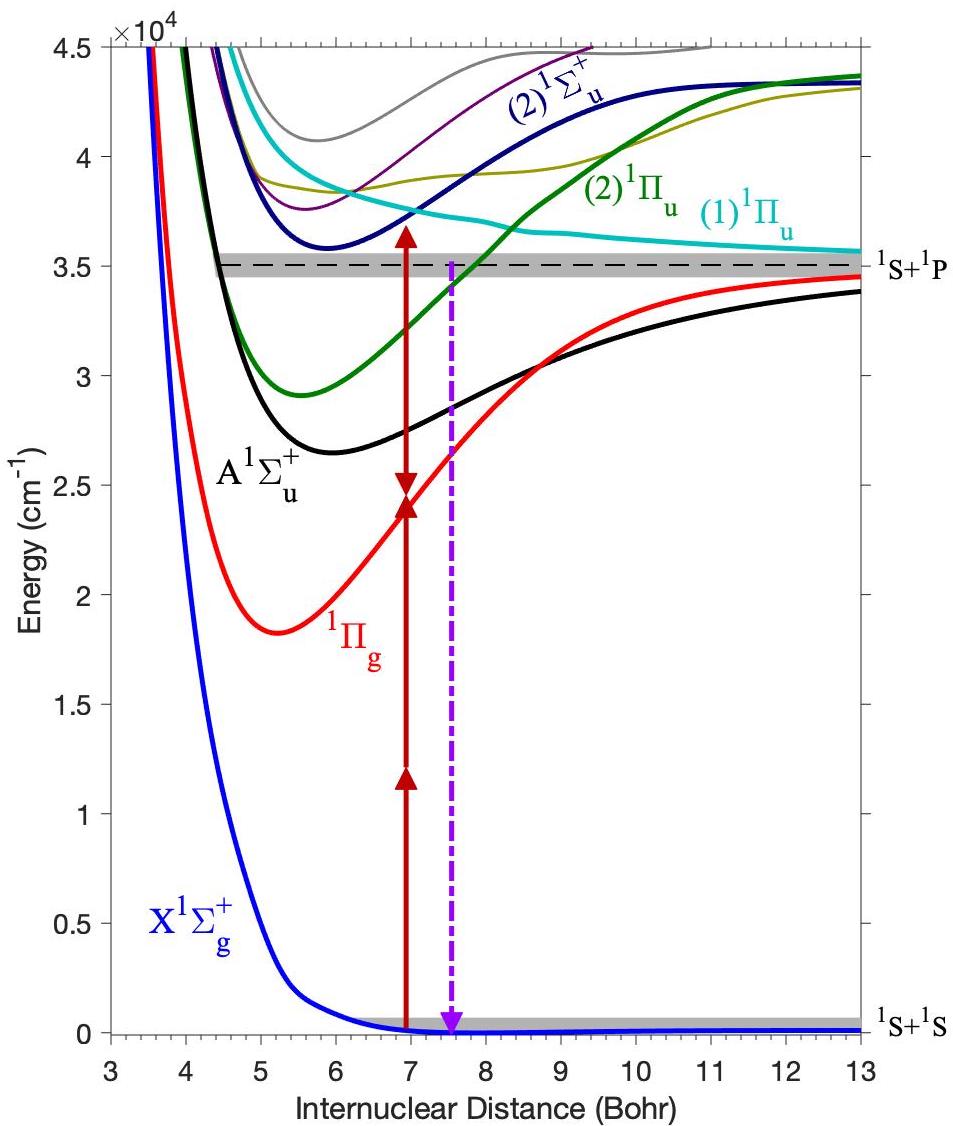}
\caption{Potential energy curves of Mg$_{2}$ and ultrafast excitation scheme corresponding to 
  an intense phase-shaped near-infrared linearly polarized femtosecond pulse.
The strong-field ultrafast excitation involves bond making (free-to-bound photoassociation) 
and subsequent molecular dynamics on the electronically excited states.
The experimental observable is the spectrum of the post-pulse UV radiation 
spontaneously emitted from the final molecular states.
The potential energy curves are plotted for the rotational quantum number $J$=70, which corresponds roughly to the maximum of the initial thermal rotational population. The gray shades indicate the initial thermal population of scattering states and the electronically excited post-pulse population which decays by UV emission.
}
\label{fig:pot}
\end{figure}
Figure~\ref{fig:pot} shows the potential energy curves for the magnesium dimer~\cite{AmaranJCP13,LevinPRL15,LevinJPhysB15} 
together with the ultrafast excitation scheme.
The starting point is an ensemble of magnesium atoms at a temperature of 1000 K 
that thermally populates scattering states above the van-der-Waals ground electronic state $X^{1}\Sigma_{g}^{+}$ of Mg$_{2}$, indicated by gray shading in Fig.~\ref{fig:pot}.
The ensemble is irradiated with shaped femtosecond laser pulses
having a central wavelength of 840~nm, a transform-limited (TL) duration of 70~fs, linear polarization, and an energy that corresponds to a TL peak intensity of 7.2$\times$10$^{12}$~W/cm$^{2}$.
The pulse photo-associates pairs of magnesium atoms and generates Mg$_{2}$ molecules 
via a free-to-bound non-resonant two-photon transition 
from  $X^{1}\Sigma_{g}^{+}$ scattering states  to rovibrational levels of the $^{1}\Pi_{g}$ state.
The corresponding Franck-Condon window is located at short internuclear distances.
Then, the pulse further induces subsequent molecular dynamics on the $^{1}\Pi_{g}$ and higher electronically excited states, 
resulting in a post-pulse population that spans a manifold of final molecular states.
%
Our present objective is to control the population branching ratio among different final states of the photoassociated Mg$_{2}$ molecules 
that cover an extended energy band located below and above the $^{1}\Pi_{g}$ asymptote, indicated also by gray shading in Fig.~\ref{fig:pot}. 
These levels belong to the $A^{1}\Sigma_{u}^{+}$, $(1)^{1}\Pi_{u}$, $(2)^{1}\Pi_{u}$ or $(2)^{1}\Sigma_{u}^{+}$ electronically excited states, 
which are all dipole-coupled to the $^{1}\Pi_{g}$ and $X^{1}\Sigma_{g}^{+}$ states. 
The $^{1}\Pi_{g}$, $A^{1}\Sigma_{u}^{+}$ and $(1)^{1}\Pi_{u}$ states share the same asymptote of Mg($^{1}P$)+Mg($^{1}S$), 
having with the ground-state asymptote Mg($^{1}S$)+Mg($^{1}S$) 
an atomic transition energy of 35051~cm$^{-1}$   
corresponding to a measured wavelength of $\lambda_{a}$=285.2~nm.    
For the control, we employ linearly-chirped femtosecond laser pulses, 
all having the same spectrum but different spectral phase.
Their spectral phase is of the form $\Phi(\omega) = \frac{1}{2} k (\omega-\omega_{0})^{2} $, where 
$\omega$ is a given frequency, 
$\omega_{0}$ the central frequency of the pulse spectrum (corresponding here to a wavelength of 840~nm) 
and $k$  the linear chirp parameter. The latter ($k$) is the control variable. 
The unchirped, or TL, pulse corresponds to $k=0$.

Experimentally, magnesium vapor with a pressure of about 5~Torr is held in a heated static cell at 1000 K with Ar buffer gas,
and the sample is irradiated at 1~kHz repetition rate with the intense shaped femtosecond pulses described above. 
They are phase shaped using a liquid-crystal spatial light modulator.~\cite{WeinerRevSciInst00}
The post-pulse population of the final molecular states is probed by detecting 
and spectrally resolving the ultraviolet (UV) radiation emitted in their spontaneous radiative decay to the ground electronic state $X^{1}\Sigma_{g}^{+}$.
The spectral range of the detected UV emission is 
$\lambda_{\text{uv},S}$=281.5~nm to $\lambda_{\text{uv},L}$=287.5~nm (34783$-$35524~cm$^{-1}$).
In the setup, the UV radiation emitted toward the laser-beam entrance to the cell is collected 
at a small angle from the laser-beam axis using an appropriate optical setup,  
spectrally resolved by a spectrometer with 0.1-nm resolution, 
and then intensity-measured using a time-gated camera system with a post-pulse gate of 20~ns.
The spectral range and resolution of the UV emission are the major differences 
to our earlier  studies~\cite{LevinPRL15,LevinJPhysB15,AmaranJCP13,RybakPRL11} 
which measured only the total UV emission intensity integrated over the range of wavelengths longer than $\lambda_{a}$. 

\begin{figure}[tbp]
\includegraphics[width=\linewidth]{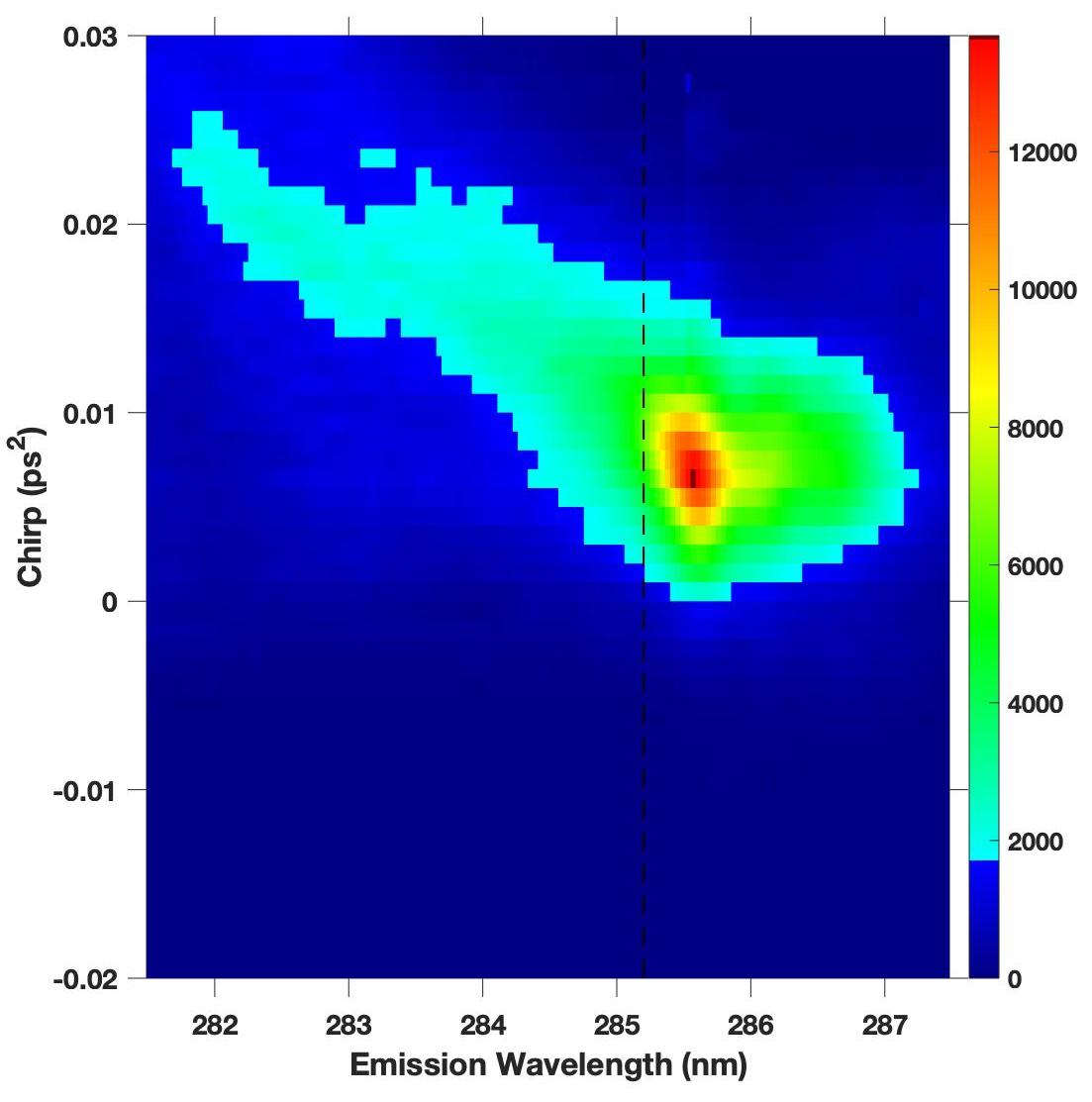}
\caption{Experimentally detected UV emission intensity as a function of emission wavelength $\lambda_{\text{uv}}$ and  chirp parameter $k$ of the linearly chirped femtosecond pulses. 
 The intensity values are color coded, and the black dashed line indicates the wavelength $\lambda_{a}$  corresponding to emission at the $^{1}\Pi_{g}$ atomic asymptote.
}
\label{fig:map}
\end{figure}
\begin{figure}[tbp]
\includegraphics[width=0.8\linewidth]{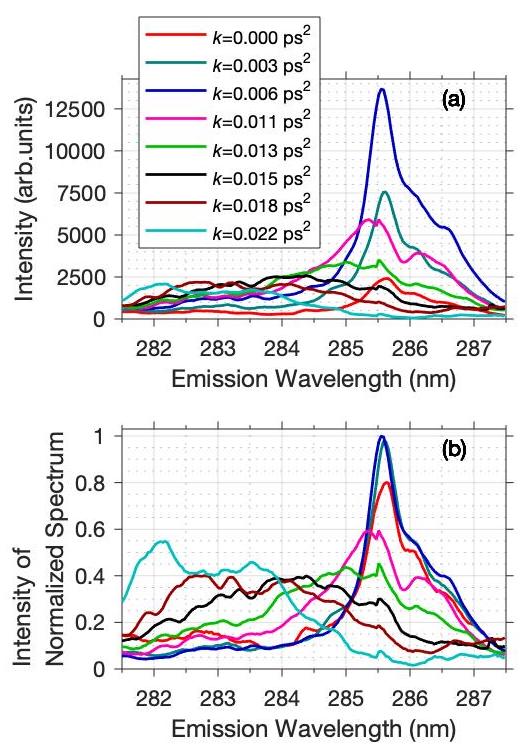}
\caption{
Examples of experimental UV emission spectra measured following the ultrafast bond-making excitation 
by linearly chirped femtosecond pulses for different values of the chirp parameter $k$. 
Each spectrum corresponds to a cut of the emission map of Fig.~\ref{fig:map} at the corresponding chirp.
For each emission spectrum, panel (a) presents the measured spectrum and panel 
(b) the spectral shape obtained by normalizing the measured spectrum by its total intensity. 
See text for details.
}
\label{fig:spectra}
\end{figure}
Figure~\ref{fig:map} shows the experimentally measured  post-pulse UV emission spectrum, obtained upon irradiation of the magnesium vapor with linearly chirped pulses. 
The results are presented as a color-coded map of the measured intensity, $I_{em}(\lambda_{\text{uv}},k)$, 
as a function of the UV emission wavelength $\lambda_{\text{uv}}$ and the chirp parameter $k$.
The  wavelength value of $\lambda_{\text{uv}} = \lambda_{a}$, corresponding to emission at the atomic transition, is indicated by a dashed black line in Fig.~\ref{fig:map}.
Out of the full map, Fig.~\ref{fig:spectra}(a) presents examples of the measured spectrum $I_{em}(\lambda_{\text{uv}} ; k)$ at several chirps. 
Figure~\ref{fig:spectra}(b) shows the same 
data normalized to the total integrated intensity $I_{em,tot}(k)$,  
i.e.,  $S_{em}(\lambda_{\text{uv}} ; k) = I_{em}(\lambda_{\text{uv}} ; k) / I_{em,tot}(k)$ with 
$I_{em,tot}(k) = \int_{\lambda_{\text{uv},S}}^{\lambda_{\text{uv},L}} I_{em}(\lambda_{\text{uv}} ; k) d\lambda_{\text{uv}} $ (with 
$\int_{\lambda_{\text{uv},S}}^{\lambda_{\text{uv},L}} S_{em}(\lambda_{\text{uv}} ; k) d\lambda_{\text{uv}}=1$  by definition).

As seen in Figs.~\ref{fig:map} and~\ref{fig:spectra}, the emission spectrum $I_{em}(\lambda_{\text{uv}} ; k)$ changes with the chirp $k$, 
such that both its shape $S_{em}(\lambda_{\text{uv}} ; k)$ and total intensity $I_{em,tot}(k)$ exhibit a strong chirp dependence.
We focus here on the change in the spectral shape.
The total emission results from a summation over the emission from the different final states, 
where each state $f$ contributes to the total measured spectrum an individual spectrum [$I_{em}^{(f)}(\lambda_{\text{uv}} ; k)$] 
with a given state-specific shape [$S_{em}^{(f)}(\lambda_{\text{uv}})$].
The total intensity [$I_{em,tot}^{(f)}(k)$] for each state is proportional to the number of Mg$_{2}$ molecules populating that state [$N^{(f)}_{mol}(k)$].
The corresponding proportionality constant [${\alpha}^{(f)}_{em}$] 
depends on the different radiative channels that the state can spontaneously decay through, 
and stands for the intensity fraction of the state's emission at the UV spectral range detected here out of the full emission from that state.
Hence, a change in the shape of the total measured spectrum [$S_{em}(\lambda_{\text{uv}} ; k)$] can result only 
from a corresponding change in the relative individual contributions of the different final states to the total spectrum.
Such a change can occur due to a change in the relative population  
[$p^{(f)}(k)$=$N^{(f)}_{mol}(k) / {\sum_{f} N^{(f)}_{mol}(k)}$] of multiple final states .
In other words, the observed chirp-dependent changes of $S_{em}(\lambda_{\text{uv}} ; k)$
directly imply corresponding chirp-dependent changes in the relative yields of photoassociated molecules populating the different final molecular states,
i.e., coherent chirp control over the corresponding yield branching ratio.

The associated quantitative analysis is as follows.
After the excitation with a pulse of chirp $k$, the emission spectrum from a final state $f$ is given by
$I_{em}^{(f)}(\lambda_{\text{uv}} ; k)$=${\alpha}^{(f)}_{em} \cdot N^{(f)}_{mol}(k) \cdot S_{em}^{(f)}(\lambda_{\text{uv}})$, 
such that the state-specific shape $S_{em}^{(f)}(\lambda_{\text{uv}})$ satisfies 
$\int_{\lambda_{\text{uv},S}}^{\lambda_{\text{uv},L}} S_{em}^{(f)}(\lambda_{\text{uv}}) d\lambda_{\text{uv}}$=1.
The chirp-dependent shape of the total spectrum emitted from a group of states is then given by 
$S_{em}(\lambda_{\text{uv}} ; k)$=$\sum_{f} W^{(f)}(k) \cdot S_{em}^{(f)}(\lambda_{\text{uv}})$
where
$W^{(f)}(k)$=
$\frac{{\alpha}^{(f)}_{em} \cdot p^{(f)}(k)} {\sum_{f} {\alpha}^{(f)}_{em} \cdot p^{(f)}(k)}$.
Both ${\alpha}^{(f)}_{em}$ and $S_{em}^{(f)}(\lambda_{\text{uv}})$ don't depend on the chirp $k$,  
while $p^{(f)}(k)$ might have such a dependence. 
Thus, 
indeed, the changes observed for $S_{em}(\lambda_{\text{uv}} ; k)$ 
when tuning the chirp $k$ necessarily reflect changes in the relative population $p^{(f)}(k)$ of multiple final states.

The chirp dependence of the measured emission spectrum
seen in Figs.~\ref{fig:map} and~\ref{fig:spectra}
includes several prominent characteristics. 
In terms of the total spectral intensity $I_{em,tot}(k)$, 
the positively-chirped pulses enhance it with respect to the TL pulse, while the negatively-chirped pulses attenuate it.
In terms of the spectral shape $S_{em}(\lambda_{\text{uv}} ; k)$, 
from chirp zero up to a positive chirp $k$ of 0.006-0.007~ps$^2$, 
it exhibits only weak chirp dependence and its dominant part is the longer-wavelength part of $\lambda_{\text{uv}}$$>$$\lambda_{a}$.
Then, as the chirp $k$ increases to larger positive values, the spectral shape shifts to shorter wavelengths, 
with an enhancement of its shorter-wavelength part and an attenuation of its longer-wavelength part.
In terms of the complementary characteristic of 
the emission intensity $I_{em}(k; \lambda_{\text{uv}})$ at a given wavelength $\lambda_{\text{uv}}$,
for all the detected wavelengths, it is enhanced 
with positively-chirped pulses over an extended 
range of positive chirps, and has a maximum at an emission wavelength-dependent positive chirp $k_{max}(\lambda_{\text{uv}})$.
The value of $k_{max}(\lambda_{\text{uv}})$ stays unchanged (as 0.006~ps$^2$) 
for $\lambda_{a}$$<$$\lambda_{\text{uv}}$$<$$\lambda_{\text{uv},L}$, 
and then continuously increases as $\lambda_{\text{uv}}$ continuously decreases from $\lambda_{a}$ down to $\lambda_{\text{uv},S}$.

Our interpretation for the mechanism facilitating the branching-ratio control is built upon the dynamics 
we have previously~\cite{LevinPRL15,LevinJPhysB15} identified to be induced by the positively chirped pulses.
Subsequent to the $X^{1}\Sigma_{g}^{+}$-to-$^{1}\Pi_{g}$ free-to-bound two-photon transition, 
the strong-field dynamics taking place on the $^{1}\Pi_{g}$ state 
involves multiple resonant Raman transitions (via  higher electronically excited states) 
that lead to a vibrational de-excitation,
i.e., a population transfer into vibrational levels that are lower than the levels accessed by the photoassociative two-photon transition.
Then, final perturbative one-photon absorption projects the vibrationally de-excited $^{1}\Pi_{g}$ population onto the final molecular states 
that emit UV light when decaying to the $X^{1}\Sigma_{g}^{+}$ state.
The higher the energetic location of the emitting states, the higher is the energetic location of the $^{1}\Pi_{g}$ vibrational region
from which they are effectively accessed by the final projection.
Following their different locations, the population transfer to distinct vibrational regions is induced by a different number of Raman transitions.
So, with a given pulse energy, each positively chirped pulse, due to its unique temporal intensity profile (i.e., a Gaussian of unique width), 
generates a different distribution of relative population among the various $^{1}\Pi_{g}$ vibrational regions 
and, thus, also among the various final emitting states.
This, in turn, results in a different 
shape for the total emission spectrum [$S_{em}(\lambda_{\text{uv}} ; k)$].
This is the control mechanism of the present chirp-dependent branching ratio among the different final states.

Furthermore, our previous results~\cite{LevinPRL15} also indicate that 
the chirp $k^{\text{(vib.region)}}_{max}$, which maximizes the population transfer into a given $^{1}\Pi_{g}$ vibrational region,   
corresponds to the positively-chirped pulse having the longest temporal segment with instantaneous intensities $J_{pulse}(t)$ that are all 
above a certain threshold intensity $J^{\text{(vib.region)}}_{th}$. The latter is associated with the vibrational region.
Increasing the chirp $k$ from zero up to this $k^{\text{(vib.region)}}_{max}$ lengthens this segment up to a maximal duration,  
and a further increase in $k$ leads to its shortening (eventually down to zero duration).
For a given pulse energy, $k^{\text{(vib.region)}}_{max}$ is inversely proportional to $J^{\text{(vib.region)}}_{th}$. 
In other words, in order to increase the strong-field de-excitation efficiency into a specific $^{1}\Pi_{g}$ vibrational region,  
it is beneficial to temporally stretch the positively-chirped pulse as much as possible up to a certain limit  (that depends on the vibrational-region's location). 
This fits the strong-field non-perturbative nature of the de-excitation dynamics.
Since 
an efficient population transfer to a higher $^{1}\Pi_{g}$ vibrational region requires a smaller number of efficient Raman transitions,  
the corresponding threshold intensity $J^{\text{(vib.region)}}_{th}$ is smaller and 
the corresponding $k^{\text{(vib.region)}}_{max}$  is larger.
As emitting states of higher excitation energy are linked with a higher vibrational region, 
their population and emitted radiation will thus be maximally enhanced 
with a larger chirp value as compared to emitting states of lower excitation energy.
Hence, the present experimental chirp dependence of the total spectral shape $S_{em}(\lambda_{\text{uv}} ; k)$ 
and the emission wavelength-dependent optimal chirp $k_{max}(\lambda_{\text{uv}})$ 
point toward the following characteristic for the states that dominantly contribute to the detected emission:
The higher the energy of the emitting state, 
the shorter are the UV wavelengths at which its state-specific emission spectral shape $S_{em}^{(f)}(\lambda_{\text{uv}})$ 
is intensified. 
Considering this characteristic in combination with the results of the theoretical emission analysis described next  
allows us to identify the dominant emitting states.

Using the potential curves presented in Fig.~\ref{fig:pot} and the corresponding electronic transition dipole moments\cite{AmaranJCP13,LevinPRL15,LevinJPhysB15},
we have assigned the origin of the experimentally detected UV emission to different electronic states and vibrational levels.   
First, we employ energetic considerations to group the different states according to their emission wavelength. 
Second, analysis of the corresponding vibrationally-averaged transition dipole moments (VTDMs) 
with the electronic ground state $X^{1}\Sigma_{g}^{+}$ allows us to 
calculate the full spectrum of spontaneous emission from each state in these groups.  
In the experimentally detected range of $\lambda_{\text{uv},S}$$<$$\lambda_{\text{uv}}$$<$$ \lambda_{\text{uv},L}$, 
four such distinct sets of states can be identified: 
(I) high-lying vibrational levels of the $A {^1\Sigma_u^+}$ state near the Mg($^{1}P$)+Mg($^{1}S$) asymptote, 
(II) vibrational levels of the two coupled $^1\Pi_u$ states below the Mg($^{1}P$)+Mg($^{1}S$) asymptote, 
(III) vibrational levels of the two coupled $^1\Pi_u$ states above the Mg($^{1}P$)+Mg($^{1}S$) asymptote 
but below the avoided crossing of the $^1\Pi_u$'s potentials,
and 
(IV) low-to-moderately high-lying vibrational levels of the $(2){^1\Sigma_u^+}$ state. 
States from groups (I) and (II) contribute exclusively to the emission at wavelengths longer than $\lambda_{a}$. 
States from groups (III) and (IV) contribute both to the emission at wavelengths longer and shorter than $\lambda_{a}$. 
Even though it would be energetically possible for even higher-lying states to add to the detected emission window 
by decaying into vibrationally high-lying levels in the electronic ground state, the VTDMs for such transitions are effectively zero. 
Moreover, vibrational levels on the two coupled ${^1 \Pi_u}$ states above the avoided crossing 
are subject predissociation which is rapid compared to the spontaneous-emission timescale. This implies that their contribution to UV emission can be neglected, and group (III) only contains levels below the avoided crossing.

\begin{figure*}[tbp]
\includegraphics[width=0.245\linewidth]{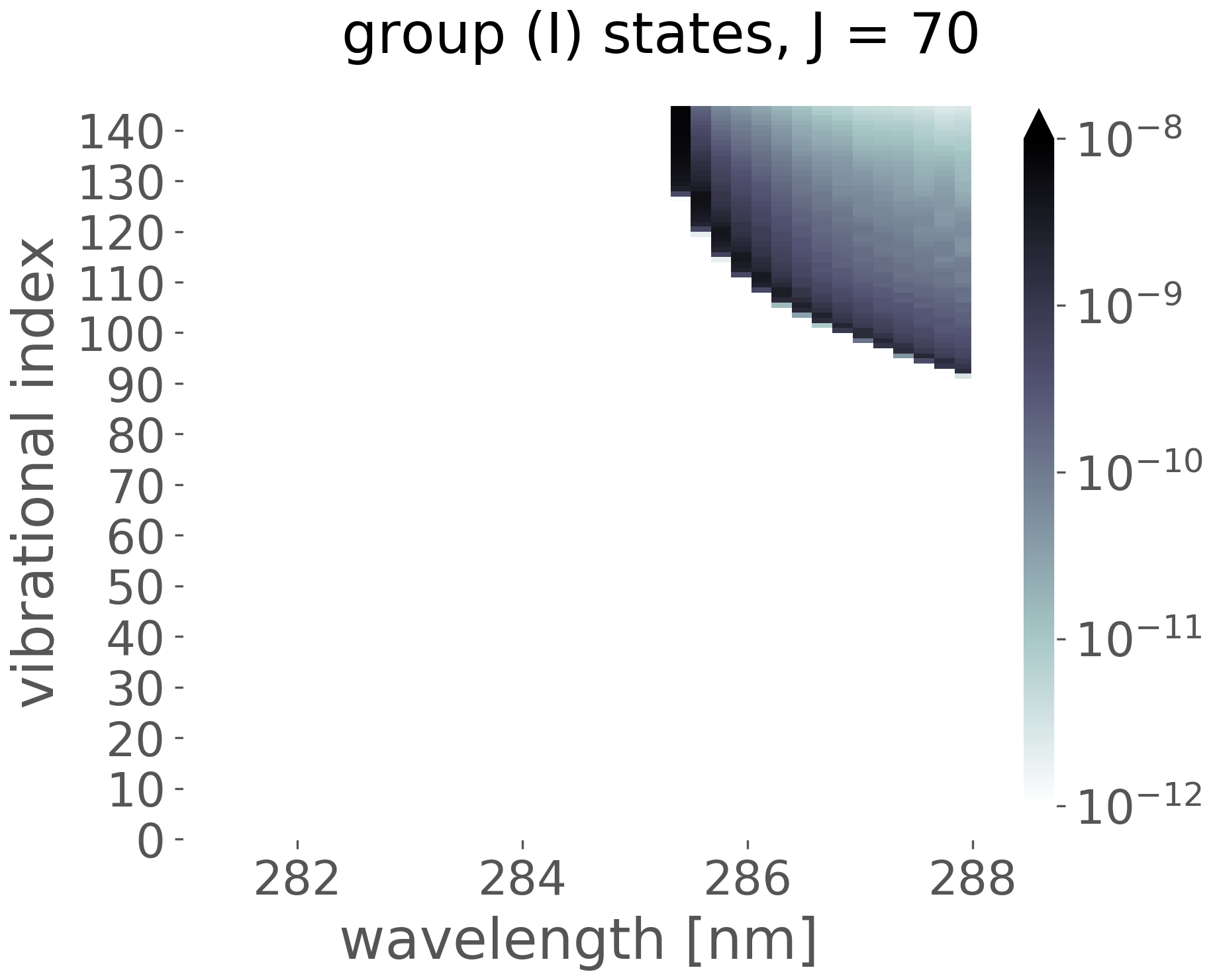}  
\includegraphics[width=0.245\linewidth]{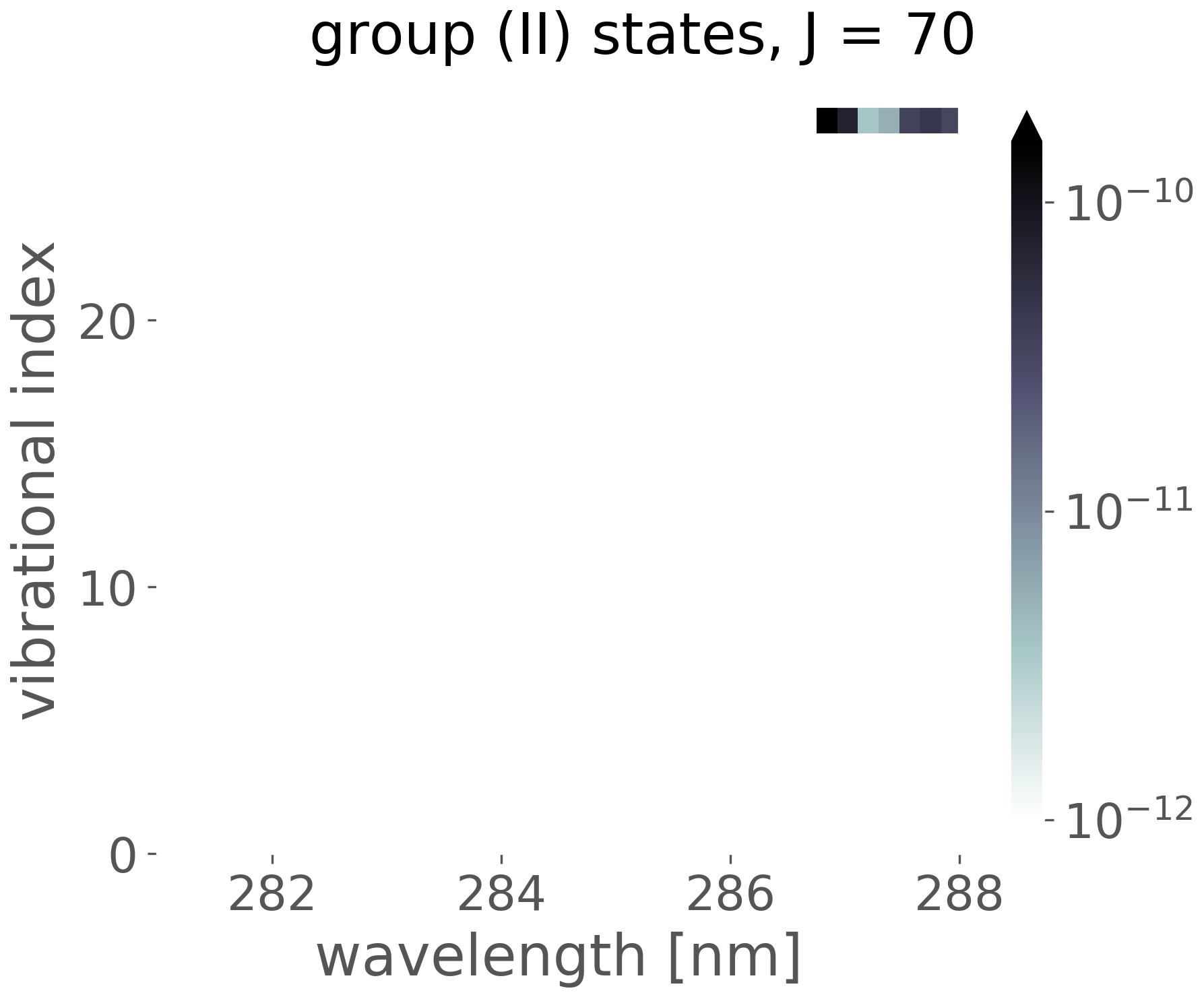}
\includegraphics[width=0.245\linewidth]{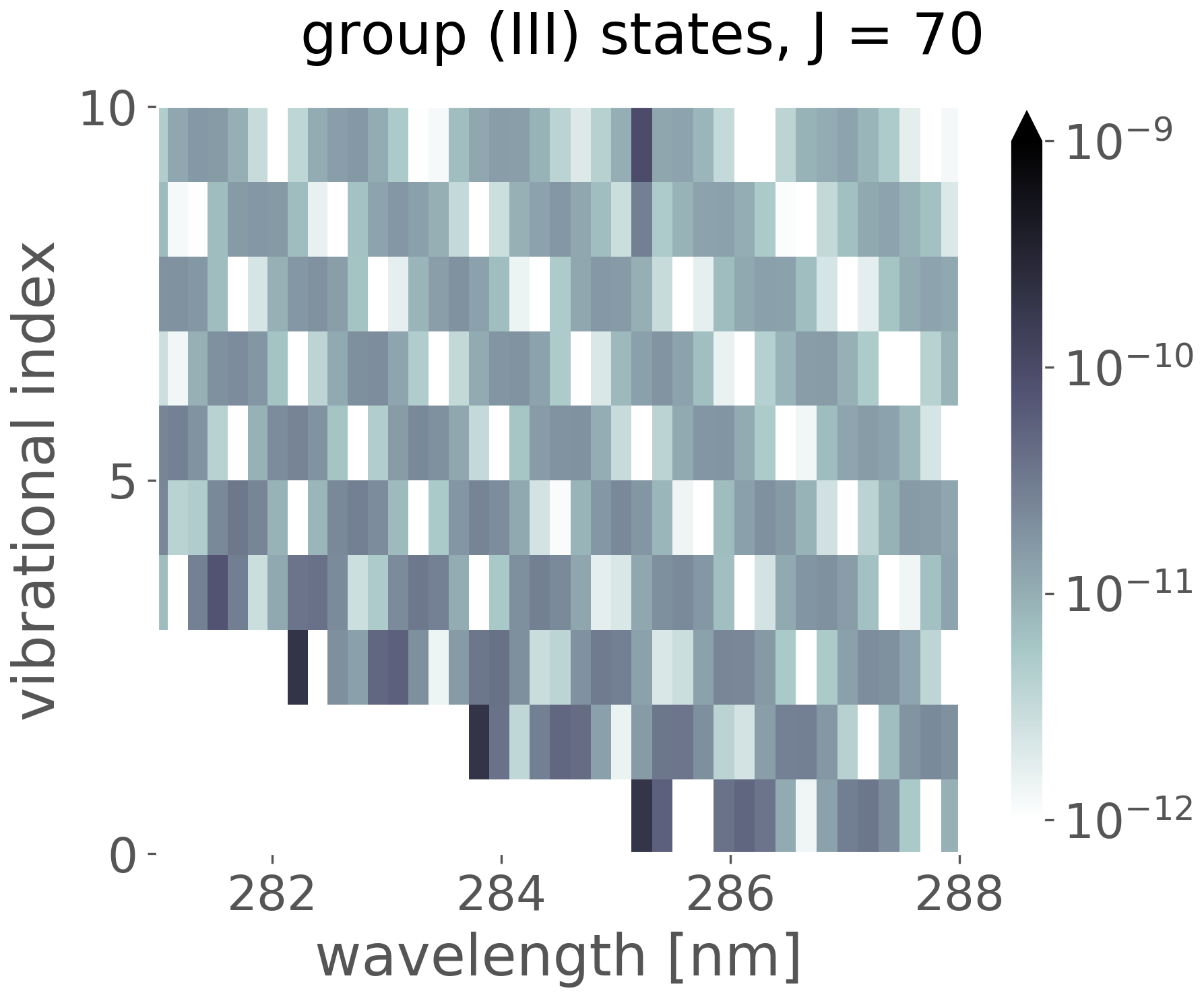}
\includegraphics[width=0.245\linewidth]{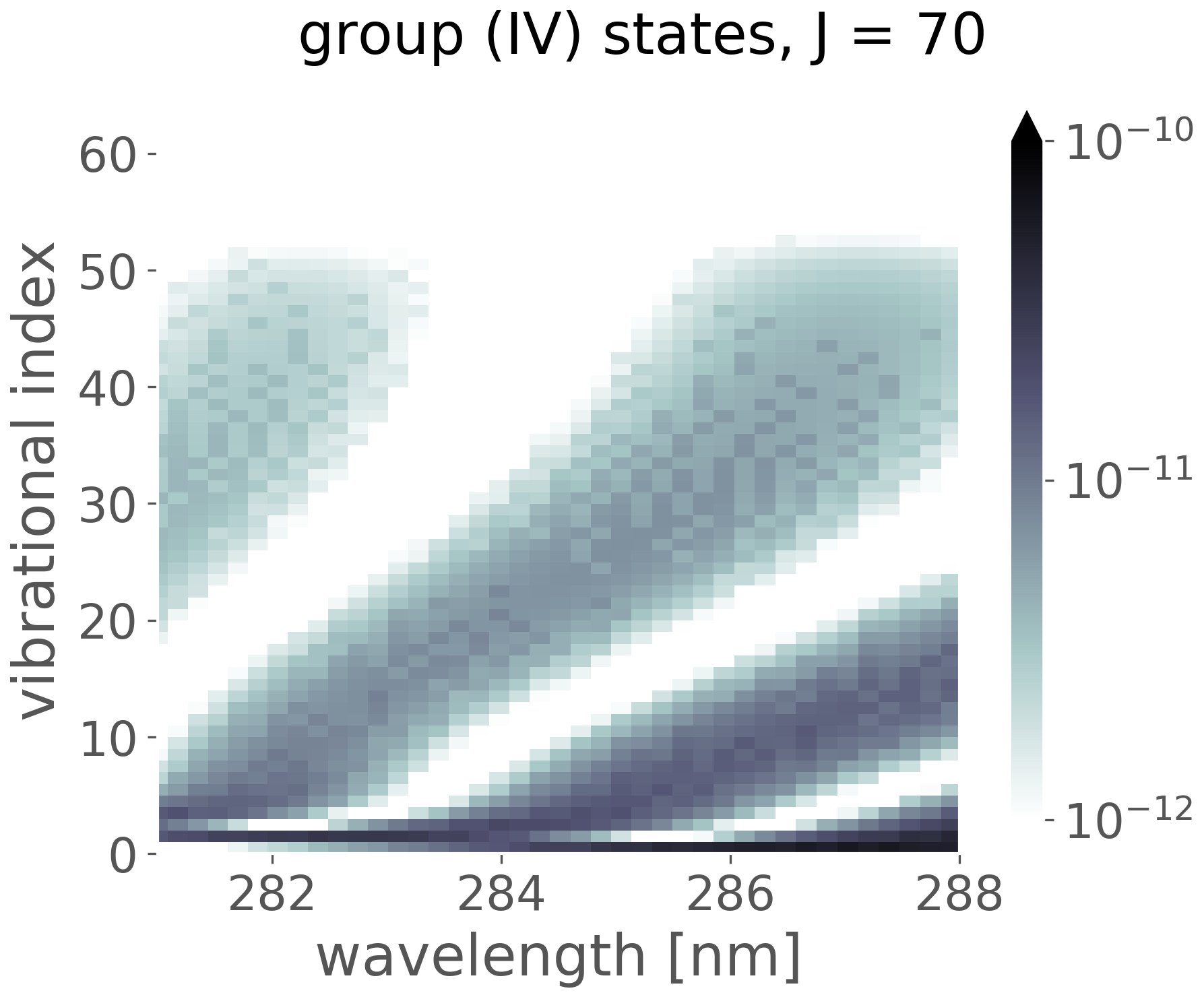}
\caption{
State-specific UV emission spectra in the experimentally detected range  
calculated for the vibrational levels with rotational quantum number $J$=70 
that belong to the four different groups of electronically excited states which may contribute to the detected emission.
Each (grayscale-coded) map corresponds to a different group, and for each vibrational level in a group, the map shows the relative emission intensity 
at each wavelength assuming the initial population of the level is 100\%. 
}
\label{fig:groups}
\end{figure*}
Figure~\ref{fig:groups} shows the group-specific UV emission spectra in the experimentally detected range,  
calculated for 
rotational quantum number $J$=70
which corresponds roughly to the maximum of the initial thermal rotational population.
The results are shown as grayscale-coded maps, with each map corresponding to a different group. 
For each vibrational level in a group, the map shows the relative emission intensity at each wavelength 
assuming the initial population of the level is 100\%.

A careful analysis of the calculated VTDMs and emission spectra for all rotational levels reveals that, for group (II), only very few vibrational  contribute to the experimental emission window. For the example of $J=70$ shown in Fig.~\ref{fig:groups}, only a single contributing level remains.
Furthermore, the group-specific emission intensities within the experimental emission window are larger for group (I) by two orders of magnitude larger compared to all other groups. 
Due to the fact that group (I) only contributes to emission at wavelengths longer than $\lambda_{a}$, 
we conclude that the detected emission at this range is dominated by group (I), 
with several dozen vibrational levels showing large VTDM values and high emission intensities. 
In particular, the density of vibrational levels in group (I) is highest right below the asymptote of Mg($^{1}P$)+Mg($^{1}S$),
which leads to an accumulation of the emission intensity directly at wavelengths slightly longer than $\lambda_{a}$.
As seen in Fig.~\ref{fig:groups}, 
the identification of group (I) as dominant in the emission spectral range of $\lambda_{a}$$<$$\lambda_{\text{uv}}$$<$$\lambda_{\text{uv},L}$  
indeed fits the characteristic identified above for the dominant emitting states 
with a correlation between  higher state energy and  intensified emission at shorter wavelengths.
Moreover, this identification of group (I) agrees with our results from previous chirp-dependent experiments and quantum dynamical calculations 
for the long-wavelength emission.~\cite{LevinPRL15} 

With regard to the emission at wavelengths shorter than $\lambda_{a}$,  Fig.~\ref{fig:groups} shows that in this spectral range only group (III) follows the aforementioned characteristic of the dominant emitting states.
Hence, we conclude that the detected emission at the spectral range of $\lambda_{\text{uv},S}$$<$$\lambda_{\text{uv}}$$<$$\lambda_{a}$ 
is dominated by the states of group (III).
This is also consistent with the larger state-specific emission intensities seen in Fig.~4 for group (III) as compared to group (IV).
Overall, we attribute the detected emission at $\lambda_{\text{uv}}$$>$$\lambda_{a}$ 
primarily to high-lying vibrational levels of the $A {^1\Sigma_u^+}$ state located near the Mg($^{1}P$)+Mg($^{1}S$) asymptote, 
whereas the detected emission at $\lambda_{\text{uv}}$$<$$\lambda_{a}$ predominantly appears to originate from 
vibrational levels of the two coupled $^1\Pi_u$ states located above the Mg($^{1}P$)+Mg($^{1}S$) asymptote 
but below the avoided crossing.


In summary, coherent control of ultrafast bond making and subsequent molecular dynamics in experiment and theory has been extended here to demonstrating branching ratio control. 
The relative yields of molecules, which are coherently generated in thermal femtosecond photoassociation and 
subsequently photo-driven into different target states, are coherently controlled using linearly-chirped pulses of positive chirp. 
The control knob is the chirp of the pulse.
The results are explained by calculations of the UV emission together with 
a model accounting for vibrational de-excitation that is photo-induced and controlled via strong-field chirped Raman transitions.
Our demonstrated control is a crucial element for the realization of coherent control over the complete process of a binary photo-reaction.
In the case of a reaction with several product channels, 
the branching ratio control to various target molecular states will be the first part of an extended scheme, 
in which the different target states will serve as intermediate platforms   
from which subsequent selective photo-control will be applied toward different product channels.

\begin{acknowledgements}
Financial support from the Deutsche Forschungsgemeinschaft (DFG – German Research Foundation) under the DFG Priority Programme 1840, 
"Quantum Dynamics in Tailored Intense Fields (QUTIF)" is gratefully acknowledged.
\end{acknowledgements}



\bibliography{mg}

\end{document}